\documentclass[prl,twocolumn]{revtex4}
\usepackage[dvips]{graphics,color}
\usepackage{amsmath}
\usepackage{amsfonts}

\usepackage{amssymb}
\usepackage{amstext}
\usepackage[sort&compress]{natbib}
\newcommand{\ignore}[1]{}

\newcommand{\ket}[1]{| #1 \rangle}
\newcommand{\bra}[1]{\langle #1 |}
\begin{document}

%%%%%%%%%%%%%%%%%%%%%%%%%%%%%%%%%%%%%%%%%%%%%%%%%%%%%%%%%%%%%%%%%%%%%%%%%%%%%%%
\title{Upper bounds on fault tolerance thresholds of noisy Clifford-based
quantum computers.}
%%%%%%%%%%%%%%%%%%%%%%%%%%%%%%%%%%%%%%%%%%%%%%%%%%%%%%%%%%%%%%%%%%%%%%%%%%%%%%%

\author{M. B. Plenio$^{1,2}$ \& S. Virmani$^{3,2,1}$}
\affiliation{$^{1}$ QOLS, Blackett Laboratory, Imperial College London,
Prince Consort Road, London SW7 2BW, UK}

\affiliation{$^{2}$ Institute for Mathematical Sciences, Imperial College
London, 53 Exhibition Road, London SW7 2PG, UK}

\affiliation{$^{3}$ STRI, University of Hertfordshire, College Lane, Hatfield, AL10 9AB}

\date{\today}

\begin{abstract}
We consider the possibility of adding noise to a quantum circuit
to make it efficiently simulatable classically. In previous works
this approach has been used to derive upper
bounds to fault tolerance thresholds - usually by identifying a
privileged resource, such as an entangling gate or a non-Clifford
operation, and then deriving the noise levels required to make it
`unprivileged'. In this work we consider extensions of this approach
where noise is added to Clifford gates too, and then `commuted'
around until it concentrates on attacking the non-Clifford resource.
While commuting noise around is not always straightforward, we find
that easy instances can be identified in popular fault tolerance
proposals, thereby enabling sharper upper bounds to be derived in
these cases. For instance we find that if we take Knill's \cite{Knill}
fault tolerance proposal together with the ability to prepare any
possible state in the $XY$ plane of the Bloch sphere, then no more
than 3.69\% error-per-gate noise is sufficient to make it
classical, and 13.71 \% of Knill's $\gamma$ noise model is
sufficient. These bounds have been derived without noise
being added to the decoding parts of the circuits. Introducing such noise
in a toy example suggests that the present
approach can be optimised further to yield tighter bounds.

\end{abstract}

\maketitle

An important open problem in the field of quantum computation
is to understand the effects of noise on computational power.
In particular, for various noise models and various sets of
universal quantum resources we would like to determine the so
called {\it fault tolerance threshold} \cite{AB}, the level
of noise that we can tolerate in our basic physical components
before we lose the power to perform quantum computation. Most
previous work on this question has focused on the problem of constructing
{\it lower bounds} to such thresholds (see e.g. \cite{AB}).
This is usually achieved by constructing explicit fault tolerant
procedures, and then {\em estimating} the level of noise that these schemes
can tolerate. The construction of {\it upper
bounds} on the other hand has received comparatively less attention, but has
recently been the subject of an increasing number of investigations.

There are essentially two
methods that have been used to derive upper bounds, which we
may loosely describe as `classical' and `quantum' approaches.
In `classical' approaches one tries to determine the noise levels
required such that the noisy quantum computer can be efficiently
simulated on a classical computer \cite{HN,VHP05,BK,Bravyi,Buhrman,ABnoisy}.
In `quantum' approaches one tries to show that above a certain
level of noise the output is bounded away from ideal in some
way \cite{ABnoisy,Razborov,Kempe,Fern,Kay}. A number of investigations
along these lines have resulted in upper bound estimates relevant
to various architectures and noise models under varying degrees
of assumption and rigour. The two approaches could ultimately be
of independent interest, as it is possible that for realistic
noise models an `intermediate' form of computation could exist
that is not as powerful as a quantum computer, but more powerful
than classical.

In this work we will follow the path of \cite{BK,VHP05,Buhrman},
and consider `classical' threshold upper bounds for a specific
class of fault tolerant quantum computational schemes - those
built around a core set of {\it Clifford} resources \cite{Nielsen C}
such as the CNOT, Pauli rotations, Hadamard, and
preparation/measurement in Pauli operator eigenbases. Any device
consisting of such Clifford operations can be efficiently simulated
classically, and so to perform quantum computation an additional
resource is needed. Common choices include phase gates of the form
\begin{equation}
        U(\theta) := |0\rangle \langle 0 | + \exp{i\theta}
        |1\rangle \langle 1|,
\end{equation}
or a supply of single qubit states such as
\begin{equation}
        |\theta\rangle = \frac{1}{\sqrt{2}}
        (|0\rangle + e^{i\theta}|1\rangle)
\end{equation}
that are not eigenstates of Pauli operators.

It is significant for our present discussion that devices built
from Clifford operations alone can be efficiently simulated
classically. This is the content of the Gottesman-Knill
theorem \cite{GK,Nielsen C}. The Gottesman-Knill theorem is
interesting as in many other respects Clifford operations exhibit
a great deal of non-classicality - for instance they can be used
to demonstrate the GHZ paradox and may generate the long-range entanglement
that is considered to be a precondition for efficient
quantum computation \cite{LindenJozsa}. Furthermore the entanglement that
Clifford operations generate is sufficient for quantum computation
- to build a full quantum computer you only need to add the
ability to perform single qubit operations to the Clifford set.

The Gottesman-Knill theorem is important for this work as it
can be used \cite{VHP05,Buhrman} to construct classical threshold
bounds using the following simple approach. Consider a Clifford
based architecture which is made universal by the addition of
a non-Clifford resource, which we refer to as $R$. If one could
determine a noise level sufficient to turn $R$ into an operation
that lies in the convex hull of Clifford operations, then this
noise level provides an {\it upper} bound to the fault tolerance
threshold for those architectures, as such a noisy device can be
efficiently simulated classically
{\footnote{In principle this approach can also be applied to
more general architectures where all resources are non-Clifford
- one can ask for the noise required to turn all the basic
components into probabilistic applications of Clifford operations.}}.
Although one would ideally
like to derive threshold bounds that are valid for a wider variety
of possible architectures than only those based upon Clifford
operations, focussing on such architectures is nevertheless quite
relevant, as most {\it lower} bounds have been derived for Clifford
based schemes due to their significance for quantum error
correction \cite{GK}.

Although a number of interesting threshold bounds can be derived
with this approach, the noise models considered in
\cite{VHP05,Buhrman}
are in fact be relatively weak because the Clifford operations are
taken to be entirely noise free. This naturally leads to the
question of whether the bounds can be improved by considering the
(often more realistic) situation in which the Clifford operations
are also subject to noise \footnote{From an experimental
point of view this is actually a natural assumption as
in many implementations of quantum gates there is little
reason to suggest that a Clifford gate should suffer
far less noise than a non-Clifford gate.}. In this work we argue that this
is indeed the case. By a straightforward modification - allowing
the Clifford operations to be noisy, and then `commuting' the noise
onto the non-Clifford operations - we argue that the previous bounds
of \cite{VHP05,Buhrman} can be improved much further
for some important families of fault tolerant quantum computation.

`Commuting' noise around a circuit is not always straightforward
as when moved through entangling gates it can lead to unmanageable error correlations between
different qubits. Hence in this article we analyse only
comparatively easy instances where these correlations can be made
to disappear or may be accounted for easily.
Fortunately, however, it turns out that the teleportation
circuit is one such `easy' instance, and so the approach works well
for the various schemes (including the high-threshold proposal of
Knill \cite{Knill}) which use teleportation as a primitive for
`injecting' non-Clifford states into the computation
to make it universal. In such
cases the problem often reduces to simple geometrical
considerations of the Bloch sphere. We expect that the
results obtained here may be improved by developing more
sophisticated techniques to tackle the correlations arising in more
general situations.

A summary of the upper bounds obtained in this paper is given
in table (\ref{summary}). In some cases they are not too
far from conjectured lower bounds for some proposed schemes.
%\mbp In remaining parts of this paper to
%make clear which bounds are rigorous upper bounds and which
%are not rigorous.\sv

\section{Basic Definitions and Noise Models}

There are a variety of noise types and models that we will
consider. In this section we define some of these noise models,
and establish some notation. We will only consider computational
models using qubits.
\begin{itemize}
\item {\bf Clifford Operations.} Consider the `Clifford group'
of unitaries, defined for any number of qubits, which is generated
by CNOTs, Hadamards, and the `$S$' gate
$S:=|0\rangle \langle 0 | + i |1\rangle \langle 1|$. Augment this
group with measurements in the Pauli $X,Y,Z$ eigenbases, and the
ability to prepare any Pauli eigenstate. Any physical operation
that can be generated by probabilistic application of these
resources (where
the probabilities can be efficiently sampled classically)will be called a {\it Clifford operation}.
\item {\bf Phase gates.} Phase gates, denoted $U(\theta)$, are
defined as:
\begin{equation}
        U(\theta) := |0\rangle \langle 0 | +
        \exp({i\theta}) |1\rangle \langle 1|.
\end{equation}
In most of the paper we abuse notation and also use $U$ to
refer to the quantum operation (on density matrices) that $U$ corresponds to.
\item{\bf Phase states.} Phase states, denoted $\ket{\theta}$ are
defined as:
\begin{equation}
        \ket{\theta} := {1 \over \sqrt{2}}\left(|0\rangle +
        \exp({i\theta}) |1\rangle \right).
\end{equation}
They lie in the plane of the Bloch sphere containing the Pauli
$X,Y$ eigenstates.
\item {\bf Probabilistic Application of a transformation Q.}
Let $Q$ be a quantum operation acting on a quantum system. $Q$
could be a unitary or a completely positive (CP) map. We define
$N^Q_t$ to be the quantum
operation ``with probability $(1-t)$ apply the identity operation
to the system and with probability $t$ apply the $Q$ operation'',
i.e.
\begin{equation}
        N^Q_t(\rho) = (1-t)\rho + t Q(\rho)
        \label{noise}
\end{equation}
Usually we will consider $Q$ to be a Pauli operation, or some
form of depolarising operation.
\item {\bf Opposite Noise.} Consider a known, given, qubit state
$\sigma$. Define $\psi^{\perp}_{\sigma}$ to be the {\it pure} state
in the polar opposite direction to $\sigma$ in the Bloch sphere.
We define $O^{\sigma}_t$ to be the single qubit operation
``with probability $(1-t)$ apply the identity operation to the
system and with probability $t$ throw it away and replace it with
the {\bf pure} state in the opposite direction to $\sigma$ in the
Bloch sphere'', i.e.
\begin{equation}
        O^{\sigma}_t(\rho) = (1-t)\rho + t \psi^{\perp}_{\sigma}
        \label{opposite}
\end{equation}
This operation will be used to construct adversarial noise upper
bounds. Note that it is a valid quantum operation as $\sigma$ is
known and fixed, and does not depend upon $\rho$.
\item {\bf Knill's noise model.} The noise model defined in the
paper by Knill \cite{Knill} is parameterized by a noise strength
$\gamma$ in the following way: a preparation of a qubit state
$\psi$ prepares the orthogonal pure state with probability
$4\gamma/15$, a measurement of the Pauli $Z$ operator is preceded
by a Pauli $X$ rotation with probability $4\gamma/15$, a measurement
of the Pauli $X$ operator is preceded by a Pauli $Z$ rotation with
probability $4\gamma/15$, a single qubit unitary is followed by one
of the Pauli $X,Y,Z$ rotations each with probability $4\gamma/15$,
a CNOT is followed by one of the 15 non-Identity Pauli products
($I \otimes X, Y \otimes I, Z \otimes X$ etc) each with probability
$\gamma/15$.
\item {\bf Independent Depolarising noise, with strength {\it t}.}
After every non-trivial unitary operation, and before each non-trivial
measurement, apply the noisy operation $N^D_t$, where $D$ is the
totally depolarising operation:
\begin{eqnarray}
    D(\rho) = {1 \over 4} \left( \rho + X\rho X + Y \rho Y +
    Z \rho Z \right)
\end{eqnarray}
\item {\bf Simultaneous Depolarising noise, with strength {\it t}.}
After each non-trivial gate acting on $k$ qubits, each possible
non-Identity Pauli product is applied to those qubits with
probability $t/(4^n-1)$. In particular this means that after
a CNOT, for example, both qubits are `jointly depolarised'
rather than independently on each wire.
\item {\bf EPG. Probabilistic Error-per-gate Noise, with strength
{\it t}.}
This is the noise model considered in many of the
more stringent and rigorous
lower bound estimates, such as \cite{Aliferis GP,Reichardt}. Each
operation (including memory steps required while waiting for
measurement outcomes elsewhere) fails with probability $t$, but
the precise manner of failure is not specified other than being
probabilistic. It can in principle be adversarial. In particular
the output of a $k$-qubit gate can undergo some correlated, joint,
error. However, correlations in the noise affecting non-interacting
qubits are not allowed in this noise model.
\end{itemize}

%\item {\bf Independent Depolarising noise, with strength {\it t}.}
%After every non-trivial unitary operation, and before each non-trivial
%measurement, apply a Pauli $X,Y,Z$ each with probability $t/3$ on each
%involved qubit wire {\it independently}. This
%    means that after a CNOT the operators $I \otimes X, X \otimes I$
%    etc. will be applied with probability $(1-t)t/3$,
%    whereas operators containing no Identity such as $X \otimes Y,
%    Z \otimes Z$ etc will be applied with probability $t^2/9$.

\section{The Basic Idea}

To illustrate the approach it is useful to begin with a specific
example analyzed in \cite{VHP05}. In that paper it was shown that
if the Clifford operations are augmented with the ability to do
single qubit phase gates and if each non-Clifford gate $U(\theta)$ in the circuit is immediately
followed by an unwanted $Z$ rotation with probability $p$, then roughly
$p \sim 14.6\%$ of noise is sufficient to make the circuit classically
tractable (this is regardless of which phase gates are used). However,
let us now consider how this result may be applied to more realistic
noise models where Clifford operations are also subject to noise. As
an initial case let us assume the following noise model: each qubit
undergoes a dephasing operation of the form $N^Z_p$ (see eq. (\ref{noise}))
after every non-trivial gate, after every qubit state initialisation,
and before every measurement.

Consider one particular qubit wire in a particular fault tolerant
circuit. If {\bf time increases from left to right}
the qubit
might be about to enter a sequence of (idealised) operations as follows:
\begin{equation}
        C \, U \, U \, C
\end{equation}
where $C$ denotes a Clifford operation (which may involve more than
one qubit, and could be a state preparation), and $U$ denotes a
non-Clifford phase gate. The real noisy version of this wire will be:
\begin{equation}
        C \, N^Z_p \, U \, N^Z_p \, U \, N^Z_p \, C \, N^Z_p
\end{equation}
By a little circuit manipulation we can reexpress this circuit as one
in which the noise is concentrated on the non-Clifford gates. The first
step is to be generous and remove the noise acting in between the two
$U$s, and on the last Clifford gate, to give (as $N^Z_p$ is itself in the convex hull of Clifford
operations, dropping any instances of it will not affect the validity
of our arguments):
\begin{equation}
        C \, N^Z_p \, U \, U \, N^Z_p \, C .
\end{equation}
Now we can replace $U \, U$ with $U'$, as the composition of two phase
rotations will just be another phase rotation. This gives
\begin{equation}
        C \, N^Z_p \, U' \, N^Z_p \, C .
\end{equation}
Hence we see that by dropping some of the noise we can reduce a circuit
of Clifford operations and phase rotations to a circuit where (apart
from the noise) every second gate on any individual wire is a Clifford
operation. Now as our noise operation $N^Z_p$ commutes with the $U$,
this equation reduces to:
\begin{equation}
        C \, U'\, (N^Z_p)^2 \, C .
\end{equation}
This is a circuit where the Clifford operations are perfect, but we
now have {\it two} lots of noise attacking the $U'$ non-Clifford gate.
In some cases we will only have one $U$ between two Clifford operations,
and in some cases we may have many $U$s of different types. We may
also have a $U$ immediately after the preparation of a fresh qubit,
or immediately before a measurement. However, in all these cases the
reduction still works. In general it is not possible to argue that
there are more than two lots of noise attacking each non-Clifford
operation, as in principle every second operation on every qubit wire
could be a non-Clifford operation.

Hence we may now ask how high $p$ must be for {\it two} lots of noise
to take an arbitrary phase rotation $U'$ into the set of Clifford
operations - this is in contrast to the calculations of \cite{VHP05}
which used only {\it one} lot of noise. Solving this problem is a
trivial extension of the methods used in \cite{VHP05}. One must solve
\begin{equation}
        (1-2p)^2 = (1-2q)
\end{equation}
for $p$, where $q$ is the minimal noise level required for a single
application of the noise. Using the precise value for $q$
found in \cite{VHP05} we find
\begin{eqnarray}
        q = {1\over 2} \left( 1 - {1\over \sqrt{2}} \right) \sim 14.64 \%.
\end{eqnarray}
Solving for $p$ hence gives
\begin{eqnarray}
        p = {1 \over 2}\left(1- \sqrt{{1 \over \sqrt{2}}}\right) = 7.955\%.
\end{eqnarray}
So this result may be summarized as follows: for an architecture based
upon Clifford operations, phase gates, and noise where each qubit
undergoes $N^Z_p$ independently after every non-trivial gate acting
upon it (or before a measurement), then a noise level of $7.955\%$ is
sufficient to make the circuit tractable classically.
%{\footnote{It
%can be shown \cite{VHP05} that for the $\pi/8$ gate (i.e. for
%$\theta = \pi/4$) that such Z noise is also the most adversarial
%independent probabilistic qubit noise.}}
Some authors prefer to use
a different parameterization of the dephasing noise model, where
instead of applying $N^Z_p$, one applies the totally dephasing
operation $\rho \rightarrow 1/2(\rho + Z\rho Z)$ with probability
${\tilde p}$. The two cases are trivially related, and so the above
bound can be reexpressed for this noise model as
\begin{eqnarray}
        {\tilde p} = 1- \sqrt{{1 \over \sqrt{2}}} = 15.9\%.
\end{eqnarray}
These improvements on the bounds of \cite{VHP05} have been possible because noise from a neighbouring
Clifford operation can be shifted onto the non-Clifford
resource. It is easy to see that this attack could lead to improvements
for any Clifford based architecture
provided that the noise can be shifted around the circuit easily. In
the example considered above, this `shifting' was possible because
the $N^Z_p$ noise commutes with any $U(\theta)$. Of course in general
the noise models won't commute with the non-Clifford resource so
straightforwardly, but as we shall see later, there are important
examples of fault tolerance schemes in the literature for which
this approach can be applied quite effectively.

\section{EPG bounds for Clifford operation and Phase gate or Phase
state schemes}

The bounds of the previous section have been derived for an adversarial
noise model which is independent on different physical qubit wires.
However, a similar approach can be used to derive a bound valid for an
adversarial error-per-gate model, where each non-trivial operation is
subject to probabilistic adversarial noise, but different physical
qubits undergoing the same multiqubit gate (e.g. a CNOT) can have
correlations in noise at that location in the circuit. The adversarial
error-per-gate model is more commonly adopted in estimates of rigorous
lower bounds, and hence is an important model to consider.

\begin{figure}[t]
        \resizebox{8.5cm}{!}{\includegraphics{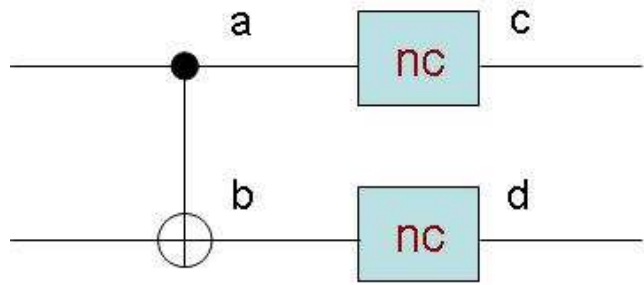}}
        \caption{A possible scenario in a fault tolerant circuit.}
\label{epgadversarial}
\end{figure}

In this case the analysis must be changed slightly. A circuit such as
that in figure (\ref{epgadversarial}) may occur, where the same Clifford
operation, in this case the CNOT, touches two non-Clifford operations.
In this case at locations $a,b$ we choose to apply the noisy operation
$N^Z_t \otimes N^Z_t$ where $t$ is chosen so that the overall probability
of any error at locations $a,b$ (of figure (\ref{epgadversarial}))
together, $t^2 + 2t(1-t)$, is equal to $p$, i.e.
\begin{equation}
        p = t^2 + 2t(1-t) \Rightarrow t = 1 - \sqrt{1-p}.
\end{equation}
Now the noise affecting each non-Clifford operation in circuit
\ref{epgadversarial} is effectively independent. As in this section
we assume the non-Clifford operations to be a phase rotation gate
(a similar argument follows through for preparation of a phase state,
with precisely the same numerical values), our goal is to work out when
\begin{equation}
        (1-2p)(1-2(1-\sqrt{1-p}))=1-2q={1 \over \sqrt{2}}
\end{equation}
which has the solution for $p$ of:
\begin{equation}
p = 10.41008383 \%.
\end{equation}
So under an adversarial EPG model, any circuit consisting of noisy
Clifford operations and phase states and phase gates can have a fault
tolerance threshold no higher than $10.41008383 \%$.

\section{Independent Depolarising thresholds for general non-Clifford
unitaries}

Depolarising noise commutes with all single qubit unitaries, so provided that the
non-Clifford operations are single qubit unitaries the above reasoning applies straightforwardly.
Although in this section we now allow
$U$ to be any non-Clifford single qubit unitary, let us
in particular take it to the most robust gate to depolarising noise,
which in \cite{Buhrman} is shown to be the $\pi/8$ phase gate anyway.
Let us ask what $p$ is required such that (where as before time increases
from left to right, $I$ denotes the identity operation, and $U$ represents the
quantum operation acting by conjugation on density matrices, not just
the unitary matrix itself)
\begin{eqnarray}
        U(\pi/4)(N^D_p)^2 &=& U(\pi/4)((1-p) I + p D)^2  \nonumber \\
        &=& (1-p)^2U(\pi/4) + (1-(1-p)^2) D \nonumber
\end{eqnarray}
is a Clifford operation (the second line of this equation follows
from the fact that $D^2=DU=UD=D$ as $D$ is the depolarising operation).
We may now apply the bound obtained in \cite{Buhrman} (a related result has been derived
independently by \cite{Bravyi}) to give that the minimal such $p$
satisfies:
\begin{eqnarray}
        (1-p)^2 = 1 - {6 - 2\sqrt{2} \over 7} = 100\% - 45.3\%
\end{eqnarray}
hence
\begin{eqnarray}
         p = 1- \sqrt{ 1 - {6 - 2\sqrt{2} \over 7}} = 26.05\%
\end{eqnarray}
This upper bound applies regardless of the single qubit non-Clifford
gates available and the fault tolerance methods used, provided that
the basic gates are Clifford + single qubit unitaries.

Ben Reichardt \cite{Ben} has independently previously performed a
related analysis for {\it simultaneous} depolarising noise. The
calculations are more difficult because in that case the noise
does not commute straightforwardly. However, it is expected that
an analysis of the simultaneous case would lead to upper bounds
of a similar magnitude \cite{Ben}.

\section{Specialising to Teleportation State-Injection schemes}

If general bounds are required for architectures involving Clifford
and non-Clifford operations, then it is generally not possible to shift more
than two lots of noise onto each non-Clifford operation. This is
because in principle every second operation in a given qubit wire
could be a non-Clifford operation.

\begin{figure}[t]
\resizebox{8.5cm}{!}{\includegraphics{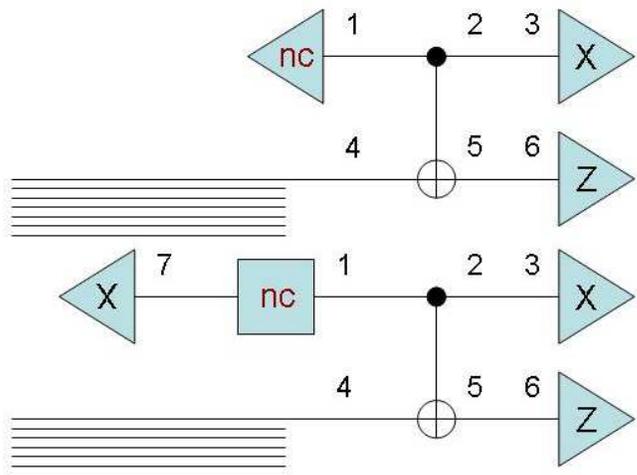}}
\caption{These circuits show two possible state injection settings.
The boxes denote gates, the triangles are either preparation (of
the +1 eigenstate) or measurement in the X,Z bases, and the collection
of wires at the bottom left of each circuit represents the postselected
decoding process. The letters `nc' denote `non-Clifford'. The top circuit
is for scenarios involving single qubit preparation as the non-Clifford
resource, the second is for single qubit gates. Each diagram has
locations numbered from 1 to 6, with the lower figure having an extra
location 7. In the text we use these numbers to define noise models
for both the upper and lower circuit in this figure. For state
injection protocols these circuits represent the only way in which
non-Clifford operations enter the computation.}
\label{injection}
\end{figure}

However, in many important proposals for fault tolerant quantum
computation the non-Clifford operations are always surrounded by
specific configurations of Clifford operations. This is because
non-Clifford operations are often introduced into the computation
using very specific ancilla preparation constructions. For example,
one method that we will consider here is the technique of
`state-injection' - see figure (\ref{injection}). This method
involves firstly creating a physical non-Clifford qubit, either by direct
access to a source of such qubits, or by applying a gate such as
the $\pi/8$ gate to a suitable Pauli-eigenstate. This qubit is
then `teleported' into an error correcting code, by first decoding
one half of an encoded bell pair to the physical level (post-selecting
on no errors), and then performing ordinary teleportation using
CNOTs and X,Z measurements. Because the teleportation circuit
immediately around the non-Clifford resource has around $5$ to $7$
non-trivial error locations (depending upon the precise model),
one can shift upto $7$ lots of noise onto the non-Clifford
resource, and obtain much lower upper bounds.
By allowing noise at these locations, and then remodelling this as
effective noise acting only upon the non-Clifford resource, it is
possible to strengthen the bounds derived in
the previous sections.

To illustrate the approach, let use consider what happens when in the
top circuit of fig.(\ref{injection}) we apply various types
of Pauli error at locations 2 to 6. The following arguments give some
rules for `shifting' these Pauli errors to location 1:
\begin{enumerate}
\item {\bf Application of $Z$ at location 6, and no errors elsewhere.}
Because location 6 is immediately followed by a $Z$ measurement,
this case is essentially equivalent to no error as it is `absorbed'
by the measurement.

\item {\bf Application of $X$ at location 6, and no errors elsewhere.}
An $X$ on the target wire commutes with the CNOT, and so the $X$ can
in fact be commuted through to location 4.

\item {\bf Application of Pauli error at location 4, and no errors
elsewhere.} Now because the CNOT and subsequent $X,Z$ measurements
implement a Bell measurement, and because each projector $B$ onto a
Bell state satisfies $(P \otimes I) B (P \otimes I) = (I \otimes P)
B (I \otimes P)$ for any Pauli operation $P$, a Pauli $P$ at location
4 is equivalent to a $P$ acting at location 1. Putting this together
with the previous point shows that an $X$ at location 6 is equivalent
to an $X$ at location 1.

\item {\bf Application of $Y$ at location 6, and no errors elsewhere.}
A $Y$ error as a {\it quantum operation} (not as a matrix) is equivalent
to a $Z$ error and an $X$ error. The $Z$ can be absorbed by the $Z$
measurement on the lower wire, leaving an $X$ error which can be moved
to location 1 by the previous points. Hence a $Y$ at location 6 is
equivalent to an $X$ at location 1.

\item {\bf Any Pauli errors at locations 4,5,6.} Because Pauli matrices
either commute or anticommute, when viewed as {\it quantum operations}
Pauli errors actually commute with each other - consider for example
the identity $XZ\rho ZX = ZX\rho XZ$. Hence the previous rules can
be applied to any set of Pauli errors acting at locations 5,6 -
effectively re-expressing them as Pauli errors acting with various
probabilities at location 1.

\item {\bf Any Pauli errors at locations 2,3.} Any $X$ errors can be
absorbed by the $X$ measurement on the top wire, any $Z$ errors can
be commuted through the CNOT to location 1.
\end{enumerate}

We consider two variants of the state injection schemes. In {\it state} resource
variants, the non-Clifford resource is a pure qubit state as in the top circuit of
figure (\ref{injection}), in {\it gate} resource variant, the non-Clifford resource
is a single qubit unitary, as in the bottom circuit of figure (\ref{injection}).

\begin{figure}[t]
\resizebox{4.5cm}{!}{\includegraphics{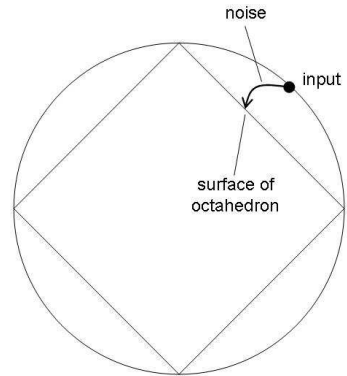}}
\caption{A 2D projection of the Bloch sphere. The goal of the calculations in this paper
is to determine the noise strength required for an input state to enter
the octahedron formed from the convex hull of the Pauli eigenstates.}
\label{vhpfig}
\end{figure}

The above rules can be applied to any configuration of Pauli noise
at locations $2$-$6$ to shift it all to location $1$, where it can
attack the non-Clifford resource. As the non-Clifford resource at
location $1$ is effectively a state in the Bloch sphere, we can solve
the relatively simple problem of how much of this noise forces the
state to enter the Clifford `octahedron' (c.f. \cite{BK,Ben magic}) formed
from the convex hull of Pauli eigenstates. In the case of upper bounds
for an EPG model, we are also free to try to pick the most adversarial
noise we can. Consider a pure Bloch vector
\begin{eqnarray} \left(\begin{array}{c}
     x \\
     y  \\
     z \end{array}\right)
\end{eqnarray}
in the positive octant of the Bloch sphere (i.e. $x,y,z \geq 0$).
A $Z$ error flips the sign of $x,y$, an $X$ error flips the sign
of $y,z$. Our goal is the find the minimal noise, for a given noise
model, which takes the input Bloch vector to an output one
\begin{eqnarray} {\rm Noise:} \left(\begin{array}{c}
     x \\
     y  \\
     z \end{array}\right) \rightarrow \left(\begin{array}{c}
     x' \\
     y'  \\
     z' \end{array}\right)
\end{eqnarray}
satisfying $x'+y'+z'=1$, which is the equation of the face of the
octahedron in the positive octant (see e.g. figure (\ref{vhpfig})).

\section{Bounds for the Knill noise model and teleportation state-injection, for
phase state and phase gate resources.}

To be perfectly clear, it is important to specify precisely how we apply Knill's noise model to
the circuits in figure \ref{injection}. In the top circuit of figure \ref{injection} we apply: a $Z$
at locations 1,3 with probability $4\gamma/15$; an $X$ at location 6 with probability $4\gamma/15$; and
at locations 2,5 considered together we apply each of the 15 non-identity
pairs of operators chosen from $I,X,Y,Z$, each pair with probability $\gamma/15$.
Note that we have kept location 4 error free. The noise at location 4 will
be determined by the decoding circuit that feeds it, and so in order to obtain general
bounds independent of the codes used we are adopting a noise model that is {\it strictly
weaker} than Knill's. Later we will discuss the effect that noise in the decoding
circuits might have.

In the bottom circuit of figure \ref{injection}, on the other hand, we apply the noise model:
at 3,7 we apply a $Z$ with probability $4\gamma/15$; at 6 an $X$ with probability $4\gamma/15$;
at 1 we apply a $X,Y$ and $Z$ each with probability $4\gamma/15$;
and at locations 2,5 considered together each of the 15 non-identity
pairs of operators chosen from $I,X,Y,Z$, each pair with probability $\gamma/15$.
Again we keep location 4 error free.

Using this noise model leads to the following effective transformation of the input Bloch vector:
\begin{eqnarray} \left(\begin{array}{c}
     x' \\
     y'  \\
     z' \end{array}\right) =  \left(1- {16 \gamma \over 15} \right)^n \left(1- {8 \gamma \over 15} \right) \left(\begin{array}{c}
     (1-8\gamma/15) x \\
     (1-8\gamma/15)^2 y  \\
     (1-8\gamma/15) z \end{array}\right) \nonumber
\end{eqnarray}
where $n=1$ for the top circuit of figure (\ref{injection}), and $n=2$ for the lower circuit of figure (\ref{injection}).
Our goal is hence to determine, for a given input resource,
the minimal $\gamma$ such that the output Bloch vector lies on the face of the octahedron.

For example, if we assume that the ideal state entering location 1 in both circuits is
$\ket{\pi/4}$, either because that is the state prepared, or because
the non-Clifford unitary is the $\pi/8$ gate, then find the solution:
\begin{eqnarray}
&& (1-16 \gamma /15)(1-8 \gamma /15)^2(1 + (1-8 \gamma /15))=\sqrt{2} \nonumber \\
 \Rightarrow && \gamma \sim 13.6861 \%.
\end{eqnarray}
for the top circuit of figure \ref{injection}, and
\begin{eqnarray}
&& (1-16 \gamma /15)^2(1-8 \gamma /15)^2(1 + (1-8 \gamma /15))=\sqrt{2} \nonumber \\
 \Rightarrow && \gamma \sim 9.5858 \%.
\end{eqnarray}
for the lower circuit of figure \ref{injection}.
Similar equations can be derived and solved for any possible phase
gate and phase state resource.
It is not difficult to solve for the minimal $\gamma$ that leads to an output vector on the face of the octahedron.
Numerically scanning through these
solutions suggests that the $\ket{\pi/4}$ and $U(\pi/4)$ resources
are not actually the most robust non-Clifford phase state or phase gate resources in this
setting, although they are very close.
If we allow {\it all} phase state and phase gate resources respectively,
the upper bounds become:
\begin{eqnarray}
\gamma \sim 13.71 \%
\end{eqnarray}
for phase states, and
\begin{eqnarray}
\gamma \sim 9.59 \%
\end{eqnarray}
for phase gates. Later we present the values obtained if any single qubit state
or gate is permitted as the non-Clifford resource.

\section{Bounds for an EPG noise model and for phase gate or phase
state teleportation state injection}

To obtain upper bounds for an adversarial EPG model, we will pick
noise which appears as detrimental as possible, yet is sufficiently
simple to analyse. It is quite likely
that the noise that we pick at each location is not the most adversarial (in the sense
of pulling the teleportation circuit into a Clifford operation) within
the EPG constraint, but this would require further analysis.

We choose the following noise for the top circuit of figure
(\ref{injection}). Assume that the input non-Clifford state is
$|\theta\rangle$. Apply $O^{\ket{\theta}\bra{\theta}}_p$ at location
$1$; $N^Z_p$ at locations 3,4; $N^X_p$ at location $6$; and replace
the CNOT with $O^{\ket{\theta}\bra{\theta}}_p$ (see eq. (\ref{opposite}))
on the top wire followed by a CNOT. This gives the following transformation
on the input Bloch vector (which has $z=0$ as it is a phase state):
\begin{eqnarray}\left(\begin{array}{c}
     x \\
     y  \end{array}\right) \rightarrow (1 - 2 p)^2 ((1 - p) (1 - 2 p) - p)\left(\begin{array}{c}
     x \\
     (1-2p)y  \end{array}\right). \nonumber
\end{eqnarray}
For an input state $\ket{\theta=\pi/4}$, the minimal
$p$ such that the output Bloch vector lies on the face of the octahedron
is:
\begin{equation}
        p = 3.68124 \%.
\end{equation}
Numerics again suggest that $\ket{\theta=\pi/4}$ is not the most robust
state in this setting, and in fact
\begin{equation}
        p = 3.69 \%
\end{equation}
is a noise level sufficient to take all possible phase states into the octahedron.

For the lower circuit of figure (\ref{injection}) we choose the
following noise. Assume that the input state is $|+\rangle$, the
$+1$ eigenstate of $X$, and that $U$ is the non-Clifford phase
gate. Apply $O^{\ket{+}\bra{+}}_p$ at location $7$;
$O^{U\ket{+}\bra{+}U^{\dag}}_p$ at location $1$; $N^Z_p$ at locations
$3$,$4$; $N^X_p$ at location $6$; and replace the CNOT with
$O^{U\ket{+}\bra{+}U^{\dag}}_p$ on the top wire followed by a
CNOT. This gives the following transformation on the output
Bloch vector (which has $z=0$ as it is a phase state):
\begin{eqnarray}
     (1 - 2 p)^2 ((1 - p) (1-4p+2p^2) - p)\left(\begin{array}{c}
     x \\
     (1-2p)y  \end{array}\right). \nonumber
\end{eqnarray}
For an input state $U(\theta=\pi/4)$ the minimal $p$ such that the
output Bloch vector lies on the face of the octahedron is:
\begin{equation}
p = 3.00339 \%.
\end{equation}
Numerics suggest that $U(\theta=\pi/4)$ is not the most robust
phase gate in this setting, and in fact
\begin{equation}
        p = 3.01 \%
\end{equation}
is a noise level sufficient to take all possible phase states into the octahedron.

\section{Bounds for teleportation injection with general Non-Clifford
states and gates}

It is not difficult to perform the analysis of the previous two sections using general
single qubit non-Clifford gate and state resources, and then to numerically
calculate the most robust of these resources. For the Knill noise
model we find that for general unitary gates
\begin{equation}
        \gamma = 15.19 \%
\end{equation}
is sufficient to turn the teleportation state-injection into a Clifford circuit, whereas
for general states
\begin{equation}
        \gamma = 21.78 \%
\end{equation}
is sufficient. The most robust non-Clifford state entering the circuit
appears to be close to, but not exactly the same as, the so-called
$|T\rangle$ state.

For an EPG model we use the same noise as for the phase gates/states
in the previous section, except at location $4$ instead of applying $N^Z_p$ we apply
$N^Y_p$. Numerical analysis of the resulting equations finds that:
\begin{equation}
        p = 5.03 \%
\end{equation}
is sufficient to turn the teleportation state-injection into a Clifford circuit, whereas
for general states
\begin{equation}
        p = 6.31 \%
\end{equation}
is sufficient.

\section{Potential effects of the decoding circuits.}

So far, we have considered solely the injection part of the circuit
as depicted in fig. \ref{injection}, and have not attempted to understand
the effects of noise in the decoding circuit. Indeed, within the framework of Knill's noise
model we have not even allowed noise at location 4.
%and have modelled the errors that
%enter the circuit at position $4$ by a depolarizing error within Knill's
%noise model. This is expected to be a reasonable first choice but
%the
The precise form of the noise generated by the decoding circuit
is difficult to determine
due to the complex structure of the encoding and decoding networks,
which may include many steps of concatenation. This brings
with it two problems.
Firstly, the entangling gates in the decoder typically generate correlations in noise which mean
that the problem becomes a multiqubit problem, rather than a simple geometrical problem on
one or two qubits. Secondly, the postselection steps have an effect on the noise
profile, and this is difficult to calculate.

However, we believe that
a careful analysis of the decoding circuits will lead to improvements in
the bounds that we have presented. In this section we present some very rough
indications of the level of improvement that might be expected, although
a more careful analysis is left to another occasion.
\begin{figure}[t]
%\hspace*{-0.75cm}
\resizebox{9cm}{!}{\includegraphics{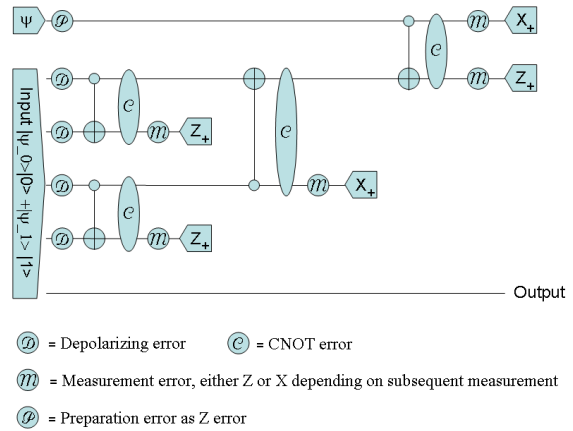}}
%\resizebox{10cm}{!}{\includegraphics{epg-adversarial}}
\caption{This circuit shows the state-injection circuit
of fig. \ref{injection} (lines 1 and 2) but now we explicitly include
the action of part of a decoding circuit with the Knill error
model. The input of the circuit is the state
$(|\psi_0\rangle|0\rangle + |\psi_1\rangle|1\rangle)/\sqrt{2}$
where the state $|\psi_i\rangle$ will be decoded by an
error free circuit to $|i\rangle$ in the second qubit.
Errors from previous stages of encoding and manipulation
are modelled by depolarizing errors.
The error free state-injection would then lead to the
preparation of the state $|\psi\rangle$ in the Output qubit
at the bottom. Determining the minimal error rate that
ensures that the Output is a stabilizer state provides an
upper bound on the fault tolerance threshold.  }
\label{encodedecode}
\end{figure}

To get a feeling for the effect of the errors that are introduced
by the decoding circuit that immediately precedes position $4$
we have carried out an analysis of the circuit in fig. \ref{encodedecode},
which corresponds to one level of decoding prior to the injection
circuit for a particular code. The input of the top arm of the
teleporter is a phase state proportional to
$\ket{0} + \exp(i\pi/4)\ket{1}$. This circuit is a simplified
version of Fig. 9 of \cite{0402171}. The simplification consists
of neglecting essentially all errors that may occur in the
preparation and encoding, as well as the decoding of the lower
half of that circuit of Fig. 9 of \cite{0402171}. We also model the error on the incoming
qubits in fig. \ref{encodedecode} via depolarizing errors (it
is in this assumption that most of the complexity is buried).

The computation of the error threshold in this approach is
possible analytically because of the small circuit size.
With the help of a computational mathematical package we find the threshold to be the sole
real root of the polynomial
\begin{eqnarray*}
        p(e_c) = -\frac{1}{2} - \frac{f_1(e_c)}{f_2(e_c)}
\end{eqnarray*}
with
\begin{eqnarray*}
        f_1(e_c) &=& (16e_c -15)^5(225 - 180e_c+32e_c^2)^2\times\\
        &\times& (16875 - 40500 e_c + 46800 e_c^2 - 23040 e_c^3 + 4096 e_c^4)\\
        f_2(e_c) &=& 1125 \sqrt{2}(576650390625 - 3536789062500 e_c \\
        &+& 11768793750000 e_c^2 - 24002325000000 e_c^3 \\
        &+& 32367600000000 e_c^4 - 29499033600000 e_c^5\\
        &+& 18141511680000 e_c^6 - 7375159296000 e_c^7 \\
        &+& 1887436800000 e_c^8 - 273804165120 e_c^9 \\
        &+& 17179869184 e_c^{10})
\end{eqnarray*}
and we find
\begin{equation}
        e_c^{threshold} = 9.2888 \%
\end{equation}
One can try to understand how critical is the choice of depolarising errors
entering the decoding steps. To estimate this we also analysed what happens if
the depolarising errors in the central four wires of figure (\ref{encodedecode}) are removed.
In this case the bound becomes
\begin{equation}
        e_c^{threshold} = 10.0638 \%
\end{equation}
Both these bounds are quite close to the bound
that would be obtained by instead modelling the noise from the decoding
circuit as a depolarising error at location 4 (for which
the bound would be 9.59\%). All these numbers should be compared
to 13.69\%, which is the number obtained for precisely the same phase state
input with no noise whatsoever outside locations 1-3 and 5-7 of Fig. \ref{injection}.

A more rigorous analysis of the decoding circuits requires much more effort and may
be reported elsewhere. However, these rough calculations give
some indication of the improvements that might be expected
in the upper bounds that we have obtained so far.

\begin{table}[t]
\caption{A summary of upper bounds presented in this work. `Injection'
refers to teleportation state-injection. `NC' means non-Clifford.}
\centering % used for centering table
\begin{tabular}{|l| l| l| l|} % centered columns (4 columns)
\hline %inserts double horizontal lines
NC resources & Method & Noise Model & Upper bound \\ [0.5ex] % inserts table
%heading
\hline % inserts single horizontal line
phase gates/states & any & independent $N^Z_p$ & 7.96 \% \\ % inserting body of the table
phase gates/states & any & EPG & 10.41\% \\
all gates & any & indep. depolarizing & 26.05\% \\
\hline
phase gates  & injection & Knill & 9.59\% \\
phase states & injection & Knill & 13.71\% \\
phase gates  & injection & EPG & 3.01\% \\
phase states & injection & EPG & 3.69\% \\
all gates  & injection & Knill & 15.19\% \\
all states & injection & Knill & 21.78\% \\
all gates  & injection & EPG & 5.03\% \\
all states & injection & EPG & 6.31\% \\ [1ex] % [1ex] adds vertical space
\hline %inserts single line
\end{tabular}
\label{summary}
\end{table}

\section{Discussion, Caveats, and Conclusion}

We have discussed `attacks' on fault tolerant quantum circuits involving Clifford
operations and extra resources, with the intention of adding the smallest
amount of noise possible to make the circuits efficiently simulatable classically.
The approach is simple - to shift noise from neighbouring Clifford gates onto
the non-Clifford resources. The approach works best for situations where this
`shifting' process can be done easily, as happens for teleportation based state injection
schemes. There are many other fault tolerance proposals in the literature (e.g.
methods built around cluster states \cite{Cluster}) that involve
a few non-Clifford resources surrounded by many Clifford operations - and so in such cases
our approach could also provide improved upper bounds, depending upon the precise
manner in which the `non-Cliffordness' is `injected' into the rest of the circuit.

We are certain that the upper bounds obtained in this work can
be optimised further, particularly through consideration of the
decoding circuits. The rigorous part of
the analysis performed in this work does not
consider the decoding circuit present in state-injection schemes,
except for a noise step at location 4 in the error-per-gate
noise model.
The analysis of the teleportation part of the state-injection is straightforward
because the teleportation circuit simplifies correlated Pauli
errors, reducing the problem to a single-qubit one.
A more sophisticated attack that also applies noise to the
decoding steps should lead to tighter bounds, especially for
Knill's noise model.

It is also important to note that teleportation is not the only method of state injection.
If the non-Clifford resource is either a measurement or a unitary, a
non-Clifford measurement may be implemented directly on the decoded wire.
This method involves far fewer error locations outside the decoder, and hence is
less susceptible to our method of attack. Hence the bounds derived in the lower
part of table (\ref{summary}) are not generally applicable to all stabilizer schemes,
but are intended more as an approach that can be fairly effective for
specific fault tolerance proposals.

A summary of the upper bounds that we have obtained is given in table (\ref{summary}).
Although the bounds that we have derived are lower than previous rigorously established upper bounds,
our bounds are not typically as general as most previous results make fewer assumptions about the architecture
(and often involve incomparable noise models).
For comparable fault tolerance schemes the estimates conjectured in [7] are
lower than the values that we have obtained, and future work will be needed to
see whether further optimisation of our approach will go as low.
However, the approach presented
here is relatively simple, makes no assumptions (other than assuming that quantum computation
is not classically tractable),
and has a slightly different aim as it deals with `classical' bounds rather than `quantum' ones.

\section{Acknowledgements}

We thank EU-STREP CORNER, the EU Integrated Project QAP, the
EPSRC QIP-IRC and the Royal Society for financial support.
We are grateful to Ben Reichardt for discussions
and for valuable comments on an earlier draft.

\end{document}